# Derivatives Discounting Explained

Wujiang Lou[1]

December 22, 2017; revised November 27, 2019

**Abstract**

Derivative pricing is about cash flow discounting at the riskfree rate. This teaching has lost its meaning post the financial crisis, due to the addition of extra value adjustments (XVA), which also made derivatives pricing and valuation a very difficult task for investors. This article recovers a properly defined discount rate that corresponds to different collateral and margin schemes. A binomial tree model is developed, enabling end-users to price in counterparty default and funding risk. Coherent XVAs, if needed, naturally result from decomposing the discount rate, and can be computed on the same tree.

Derivative pricing to is about discounting, once the underlying market dynamics is prescribed. In the classic Black-Scholes (B-S) option pricing model, the expected option payoff is discounted at the riskfree rate to arrive at the fair price. In fact, the discovery of this celebrated model was keyed on the correct discount rate, according to Black and Scholes (1973),

*"What they fail to pursue is the fact that in equilibrium, the expected return on such a hedged position must be equal to the return on a riskless asset. What we show below is that this equilibrium condition can be used to derive a theoretical valuation formula."*

The return on a riskless asset, or the riskfree rate, later became the cornerstone of the risk-neutral pricing theory and routine derivatives valuation practice. In the US, Treasury securities' zero coupon rates were historically taken as the riskfree rates since the government bonds are considered default-free (at least in their own currency) and having zero coupon removes any reinvestment risk. While this satisfied theorists, practitioners soon found a disconnect between this theoretical rate and the operational funding rate, the rate at which they got charged for their intra-bank borrowing in relation to derivatives trading. Recognizing that the treasury rate was only accessible to the government's financing, some firms attempted using Treasuries repo rates as the riskfree rates. Most of the dealer-banks were funded at LIBOR (London Intra-Bank Offered Rates) pre-crisis, so the LIBOR curve, a series of zero rates inferred from the LIBOR rates and LIBOR indexed swap rates, emerged as the *de facto* standard for derivatives and structured products.

Discounting on the LIBOR curve or LIBOR discounting, however, is not the complete story even prior to the financial crisis. Since non-exchange traded, over-the-counter (OTC) derivatives are traded between defaultable counterparties, they are exposed to counterparty's credit risk[2]. More sophisticated dealer banks started to compute a counterparty credit value adjustment (CVA) made to the default risk free fair price. Consequentially the riskfree rate is no longer the discount rate leading to the final price.

---

[2] Since LIBOR is an average of intra-bank unsecured rates, its replacement of the treasury curve as the riskfree rate is thought to have largely taken into account of intra-bank counterparty credit risk. Some consider this as the justification to use LIBOR, Huge and Lando (1999).

The fact that CVA definitions vary and often get watered down in implementations makes difficult to perceive that a derivative discount rate even exists.

The surge in TED (Treasuries and Euro Dollar deposit, i.e., Treasuries and LIBOR) spread during the credit crunch of 2007-2008 readily rejected LIBOR as a good approximation of the riskfree rate, as the difference of the three month LIBOR rate and Treasuries bill rate spiked, for instance, to 4.64% on 10//10/2008. In the meanwhile, banks were found unable to fund anywhere close to LIBOR levels, indicating that banks' credit risk could diverge from what LIBOR represents. The practice of discounting derivatives cash flows exclusively at the LIBOR is thus no longer appropriate, and LIBOR itself has been subject to intense scrutiny and pending replacements[3].

Realizing that fully collateralized derivatives are essentially counterparty credit risk free and cash funded at the overnight indexed swaps (OIS) rates, the industry quickly forged a consensus to discount such derivatives at the OIS curve, or OIS discounting. The rest of the OTC derivatives were initially left with LIBOR discounting, or the same OIS discounting (Hull and White 2013). This didn't last long however, as some banks, aware of the fact they funded away from LIBOR, started using their own funding rates to compute a funding valuation adjustment (FVA) on uncollateralized derivatives.

As Hull and White (2014) pointed out, this FVA equates to discounting an investment's cash flows at the investor's funding rate rather than the project's risk based return rate, and as such is at conflict with fundamental principles taught in finance. Typical FVA definitions, however, no longer allow a clear identification of a discount rate applicable to future derivatives cash flows. The industry has since endeavored to mitigate counterparty risk by moving towards to central counterparty (CCP) clearing and enforcing similar margin and collateralization schemes if non-CCP cleared. Other value adjustments (XVA) emerged, such as MVA for margin, and LVA for collateral liquidity, complicating the matter further. It is safe to say that no trace of a unique discount rate can be found in the final price, when all these XVAs are accounted for.

---

[3] LIBOR Sunset: LIBOR is set to discontinue in the last quarter of 2021, to be replaced by the Secured Overnight Financing Rate (SOFR). It includes tri-party Treasuries repo data, and cleared bilateral and GCF repo data from the Depository Trust & Clearing Corp.

On the collateral side, different types of assets and margining schemes are possible. Cash collateral, for instance, could be segregated, treasury securities are commonly admitted in lieu of cash, and non-government securities are often accepted. The primary purpose of derivatives collateral is of course to mitigate counterparty credit risk, but it also serves a secondary purpose of funding. Non-segregated cash collateral, for example, can be used by the receiving firm for its general corporate purposes. It can certainly be used to fund the dealer's operations, and thus brings funding benefit as its interest rate is the lowest possible, the overnight Federal Funds rate. If cash is segregated, then it can only be drawn when counterparty defaults. Segregated cash, therefore, provides only credit protection but not funding. Different collateralization schemes thus have side effects and a discount rate, if exists, would have to be commensurate.

The notion of riskfree discounting has been blurred amid serious challenges from the 2007-2009 global financial crisis. Needless to say, it has caused confusion and complicated derivatives pricing and valuation tasks, especially for derivatives end-users. This paper sets out to reestablish the discounting rate for post-crisis OTC derivatives pricing and valuation. The derived final price incorporates counterparty risk and reflects collateralization schemes. Presentation of such a rate benefits market participants by affirming our usual textbook understanding of discounting an investment project's future cash flows corresponding to the project's risk characteristics. Derivatives end-users who are familiar with binomial tree pricing models can simply modify their models to solve for the final price by properly choosing its local discount rate at each tree node during the rollback discounting step.

The same binomial tree procedure can be used to compute fair value and XVAs simultaneously, allowing end-users to come up with their own XVA estimations independently and to better understand broker-dealers or banks' OTC derivatives pricing adjustments. In the US, for example, the Dodd Frank Act requires that swap dealers provide their customers with pre-trade mid-market marks (PTMM). Their actual pricing will be shown as a spread to PTMM that could include credit, hedging, funding, liquidity, and other adjustments. Dealers' practice could vary widely, as there is no consensus whether some XVAs should exist and how they should be computed. In this paper, all these adjustments are treated as a coherent decomposition of a fair value which is unequivocally linked to components of a risk-sensitive discount rate.

**1. Incorporating stock financing in the B-S model**

Before we proceed to derive discount rates appropriate to various derivatives collateralization schemes, it is important to recognize that the riskfree rate, although critical to Black and Scholes (1973)'s discovery, is not strictly necessary. Starting with a long stock and short call option hedged portfolio, Black and Scholes argued that the hedge ratio could change so that at any given moment the hedge portfolio value change does not depend on the stock price and is thus free of market risk. Such a portfolio then can't be arbitraged against a riskless security. So in market equilibrium, "the expected return on such a hedged position must be equal to the return on a riskless asset" by virtue of the capital asset pricing model (CAPM). The option price $V(S, t)$ as a function of stock price $S$ and time $t$ is governed by the following partial differential equation (PDE) -- the Black-Scholes equation,

$$\frac{\partial V}{\partial t} + (r-q)S\frac{\partial V}{\partial S} + \tfrac{1}{2}\sigma^2 S^2 \frac{\partial^2 V}{\partial S^2} - rV = 0 \tag{1}$$

where $\sigma$ is the volatility of the stock, $q$ the stock dividend yield, and $r$ is the rate of return on a riskless asset or the riskfree rate. There are two appearances of $r$, one in the second term relating to the stock, and a second one in the fourth term to indicate that the option value is discounted at $r$. The latter is obvious in the risk-neutral pricing formula, e.g., that of a European call option with strike $K$,

$$V(S,t) = E_t^Q[e^{-r(T-t)}(S_T - K)^+]. \tag{2}$$

where $E$ denotes the usual expectation operator in the risk-neutral world $Q$ in which the expected return of the stock is the riskfree rate, $e^{-r(T-t)}$ the riskfree discount factor, $(S_T - K)^+$ call option payoff.

With regard to the first appearance, $r$ can be taken as the assumed stock financing rate. This is indeed one of the basic assumptions, assumption (f) from Black and Scholes (1973): *"It is possible to borrow any fraction of the price of a security to buy it or hold it, at the short-term interest rate, which is known and constant through time."* Black and Scholes' derivation and Merton's derivation, however, do not make explicit use of this riskfree rate financing assumption to show the funding cost is *actually* zero. $r$ in equation (1) is a result of a perfectly hedged portfolio having to return at the same rate as a riskless asset to avoid arbitrage or to achieve market equilibrium. This blanket styled funding assumption is obviously unrealistic, but it conveniently allows them to focus on finding a no-arbitrage solution to option pricing. Facing the operation aspect of hedging and financing the stock hedges, the industry has long recognized the fact that stocks can't be financed at the riskfree rate and has replaced it with a stock repo rate in practice. As shown below, assumption (f) can be made more precise and realistic by allowing stocks to be financed through the stock financing market.

To borrow money to buy stock, the cheapest way is to borrow from the repo market. A repo is a repurchase agreement where a security is sold to a financer and will be purchased back at a stated higher price on a later date. The initial purchase price is essentially a cash loan amount and the difference between purchase and repurchase prices is the financing interest. A repo is economically a loan collateralized with the bought stock shares. The loan amount is tied up to the market value of the stock shares, after a discount. For example 500 shares of stock with market price at 100 will receive a loan of 40,000 after a 20% haircut is applied.

Assumption (g) of Black and Scholes (1973) further states *"no penalties to short selling"*, which can be understood to mean that there is no stock borrowing cost and the short sales proceeds are immediately available to the short seller. Borrowing stock to short is done through the securities lending (sec lending) market. The short seller typically pays a borrowing cost to the securities lender (sec lender) for the use of the security. When the borrowed security is sold in the open market, the sales proceeds are handed over to the security lender as collateral to mitigate the borrower's credit risk. Not only that, additional cash margin is required to secure the sec lender in case of a market surge causing the borrower unable to buy back the security to return to the lender. This margin is similar to repo haircut in purposes and mechanism.

Following Merton (1973) and Shreve (2004), we set up an economy or portfolio that consists of the short European call option fair valued at $V_t$, long $\Delta_t$ shares of the underlying stock $S_t$, and a money market or bank account[4] with balance $M_t$. The portfolio value denoted by $\pi_t$ is $\pi_t = M_t + \Delta_t S_t - V_t$. Merton starts with $\pi_0 = 0$ and states that the no-arbitrage condition would require that the portfolio remain zero valued, i.e., $\pi_t = 0$ for all $t<T$, which leads to $M_t = V_t - \Delta_t S_t$. In particular, $M_0 = V_0 - \Delta_0 S_0$. For a call option, $M_0<0$, and in fact $M_t<0$. A negative balance means that we need to borrow money to fund the initial portfolio.

Riskfree borrowing of cash is of course never possible from a funding perspective. Suppose that stock can be financed at zero haircut, at a rate of $r_s$. We now expand the economy to include a repo borrow account $L_t$, with total portfolio value

$$\pi_t = M_t + \Delta_t S_t - V_t - L_t. \tag{3}$$

Because of zero haircut, we are able to borrow at the full amount, i.e., $L_t = \Delta_t S_t$. Again, the economy starts at zero value, $\pi_0 = 0$, so $M_0 = V_0$, the initial balance of the bank account is exactly the option sale proceeds. Setting $\pi_t = 0$ for all $t<T$ leads to $M_t = V_t >0$. Now $M_t$ is positive, a sure deposit, so we don't need to borrow money in this economy.

---

[4] In terms of presentation and notation, Merton (1973) uses a riskless bond, Shreve (2004) uses a money market fund account, and others use a bank (savings) account. Here we adopt the bank account notation to distinguish cash deposit from cash borrowing, as different interest rates apply. Obviously the money market fund is assumed riskless, although the late financial crisis has shown the contrary. The constant riskfree rate in the original B-S model was extended to stochastic riskfree rate by Merton (1973) who also provides an alternative derivation in continuous time showing that the B-S model can be obtained under a no-arbitrage condition, weaker than a general market equilibrium assumption under the CAPM. Merton compliments the hedging portfolio with a riskless bond. The riskfree nature of the discount rate maintains.

With initial cash flow and funding set, we consider incremental funding. Since the stock financing account is tied up to the delta hedging strategy via $L_t = \Delta_t S_t$, there will be an incremental repo funding amount of $dL_t$ available when the stock price moves in time. To rebalance delta hedging, we need to buy $d\Delta_t$ more shares of stock at the price of $S_t + dS_t$, so the payment amount is $d\Delta_t(S_t + dS_t)$. Other cash flow items include stock dividend income of $q\Delta_t S_t dt$, interest income from the bank account $rM_t dt$, and interest payment to the stock repo account $r_s L_t dt$. Now that the economy is segregated, the net cash flow in this $dt$ period is saved back into the bank account, resulting in the financing equation as follows,

$$dM_t = rM_t dt - d\Delta_t(S_t + dS_t) + q\Delta_t S_t dt + dL_t - r_s L_t dt, \qquad (4.a)$$

The last term of (4.a) is the repo interest the economy pays to the repo lender. As $dL_t = d(\Delta_t S_t) = d\Delta_t(S_t + dS_t) + \Delta_t dS_t$, the increased stock funding and payment for additional hedge purchases net out to $\Delta_t dS_t$. Therefore, apart from financing the stock purchases, the repo account may have extra capacity ($\Delta_t dS_t$) funding the economy when stock price goes up, or take away funding from the economy when $S_t$ decreases. This observation is very important as it explains how the option is funded a bit later. Rearrange (4.a) as

$$dM_t - rM_t dt = \Delta_t dS_t - (r_s - q)\Delta_t S_t dt, \qquad (4.b)$$

This is more familiar form of the self-financing equation, e.g., in Shreve (2004). $\Delta_t dS_t$ is exactly the unrealized profit-n-loss (pnl) on holding $\Delta_t$ shares. Alternatively we can think of this as realized pnl since in the B-S model, trading can be done continuously without any transaction cost. Such is the case, the realized pnl is immediately reinvested in the same stock, and it does not results cashflow. Therefore, interpreting $\Delta_t dS_t$ as realized or unrealized pnl does not make any difference. The right hand side of (4.b) is the net trading gain, trading pnl net of stock financing cost and dividend income.

From equation (4.b), replace $M_t$ with $V_t$ as $M_t = V_t$,

$$dV_t - rV_t dt = \Delta_t (dS_t - (r_s - q)S dt) \tag{5}$$

Assuming $V$ as a function of stock price $S$ and time $t$, applying Ito's lemma

$$dV_t = \frac{\partial V}{\partial t} dt + \frac{\partial V}{\partial S} dS + \tfrac{1}{2}\sigma^2 S^2 \frac{\partial^2 V}{\partial S^2} dt \text{ leads to}$$

$$(\frac{\partial V}{\partial t} + \tfrac{1}{2}\sigma^2 S^2 \frac{\partial^2 V}{\partial S^2} - rV_t)dt + \frac{\partial V}{\partial S} dS = \Delta_t dS_t - \Delta_t (r_s - q)S dt \tag{6}$$

Choosing the delta hedging strategy $\Delta = \frac{\partial V}{\partial S}$ to eliminate $dS_t$ term, equation (6) results in the B-S equation,

$$\frac{\partial V}{\partial t} + (r_s - q)S \frac{\partial V}{\partial S} + \tfrac{1}{2}\sigma^2 S^2 \frac{\partial^2 V}{\partial S^2} - rV_t = 0 \tag{7}$$

The only difference from the original B-S PDE (1) is that the stock financing rate $r_s$ has taken the place of $r$ in the second term. The second appearance of $r$ in (1) remains, but we could have used a different notation, as it now stands for the interest rate earned in the bank deposit account. Its meaning of being a riskfree rate is not necessarily lost (yet) because the cash deposit in a fed reserve bank account is both credit and market risk free.

As a relevant note, the risk neutral rate of return of the stock is $r_s$-$q$, different from the classic B-S model. Of course, this can be thought of an adjusted dividend, i.e., $q` = q - (r_s$-$r)$, so that $r_s$–$q = r - q`$. The interpretation is clear in that the stock borrowing cost has to be deducted from the stock dividend.

**1.1. Stock financing rate in practice**

As shown above, the rather broad riskfree borrowing assumption (f) of the B-S model can be narrowed down to zero haircut stock financing at the repo market rate. This pulls the B-S model closer to reality, although zero haircut is not typical in the repo market.

Practically, recognizing that haircut is usually a small percentage, we can incorporate its impact approximately. For example, traders typically treat $r_s$ as the overall funding cost, by using a simple formula $r_s = h*r_b+(1-h)*r_p$, where $h$ is the haircut, for example 0.15 for 15%, $r_p$ is the repo rate and $r_b$ is the rate applicable to the haircut portion, typically an average of dealers' unsecured financing rates.

Lou (2015a) finds that having to fund the haircut portion at unsecured rates plays a role in options market maker's bid/ask spread. To hedge a net short delta inventory, for example, a market maker needs to buy stock with repo financing and pays the unsecured rate on the haircut portion not covered in the repo market. To compensate for the funding cost, he will lower his bid. Similarly with a net long delta inventory, he needs to short stock. The additional cash margin on top of the short sales proceeds still needs to be financed at the unsecured rate. The cost so incurred pushes his ask higher. With either long or short delta inventories, the additional funding costs widen the bid/ask spread. It is shown that reasonable funding cost parameters can easily produce same magnitude of bid/ask spread of less liquid, longer term options as observed in the market place.

## 2. Derivatives Financing

In terms of financing the call option's only cash flow, i.e., the payoff at expiry $T$, with the aid of a zero-haircut stock financing market, the initial transaction price $V_0$ alone is sufficient to fund the European option's cash flow at the expiry: if stock price at $T$ is greater than strike $K$, the bank account balance would have grown to $(S_T-K)^+$, and the repo account closes with the exchange of 100 shares and money; otherwise, the money in the bank account would have been drawn to pay the repo lender and reduced to zero. This can be seen by integrating equation (4.b) over time interval *[0, T]*, and noting $M_0 = V_0$, and $M_T = V_T$,

$$d(e^{-\int_0^t r du} M_t) = \Delta_t (dS_t - (r_s - q)S_t dt) e^{-\int_0^t r du}$$

$$V_T = V_0 e^{\int_0^T r du} + \int_0^T e^{\int_t^T r du} \Delta_t (dS_t - (r_s - q)S_t dt)$$

The above equation shows that the option payoff is replicated by the initial cash deposit in the bank account and trading gain due to the delta hedging strategy. All funding needs are met and the delta hedged option economy with a zero haircut stock financing market is complete and self-financed. No riskfree borrowing needs to be assumed.

Some investors could ask whether there is a capacity of leveraging up on the option purchase. This brings about the topic of derivatives financing. Unlike stock financing that has unambiguous meaning, derivatives financing is a term warranting clarification. To be sure, there is no securities financing market that takes in derivatives as collateral to lend money or lends derivatives in the same way as stock is lent to facilitate a short. Sometimes, the term is loosely used to mean derivatives based securities financing, for instance, lending money out in the form of a total return swap (TRS) or credit default swap (CDS), instead of a repo. This usage situation is similar to the term "swap financing", which means using swaps to exchange fixed rate financing for floating rate or vice versa, or to exchange one currency for another, but has no bearing with using swaps as collateral to borrow money.

Suppose, hypothetically, a derivatives financing market exists and operates in the same way as a stock financing market. Let $r_d$ denote the derivatives repo rate. Consider the dealer-bank buys rather than sells an option as in the last section. If we treat equation (3)'s $V_t$ as algebraic, positive meaning the short position value, negative long position, then $M_t = V_t$ still stands, although $M_t = -|V_t| < 0$ now. A negative $M_t$ means the economy is borrowing from, not depositing into, the bank account. The overnight Fed Funds rate no longer applies. So the dealer-bank opens a repo transaction pledging out the option for money at the interest rate of $r_d$. By following the same derivation steps, equation (7) remains valid, except with $r$ replaced by $r_d$, and the call option's fair price is also updated,

$$\frac{\partial V}{\partial t} + (r_s - q)S\frac{\partial V}{\partial S} + \tfrac{1}{2}\sigma^2 S^2 \frac{\partial^2 V}{\partial S^2} - r_d V_t = 0 \tag{8.a}$$

$$V(S,t) = E_t^Q [e^{-r_d(T-t)}(S_T - K)^+]. \tag{8.b}$$

Comparing with pricing formula (2), the hypothetic derivatives repo financing rate becomes the discount rate.

Although hypothetic, separation of financing underlying stock hedging activity and financing the option allows the riskfree rate assumption or any of its implication to be detached from the option pricing formula.

The rest of the paper discusses a limited yet realistic form of derivatives financing and its impact on derivatives pricing through an effective derivatives financing rate or the discount rate. In this regard, we use the term derivatives financing to mean any realization of liquidity, aka cash, related to a derivatives position or trade but not to the derivative product itself. Swap margin movements, for example, are related to the trade, but are not considered to be part of the swap's native cash flows. Accordingly, the term "derivative financing rate" means the rate at which the derivative attracts liquidity, for instance, the OIS rate for fully cash collateralized trades.

## 2.1. Financing in fully cash collateralized derivatives

A cash collateralized derivative entails a daily exchange of cash margins in the amount of derivatives' market-to-market, or fair value changes. It essentially sets up a private settlement mechanism same as exchange traded futures contracts.

Consider the ATM call option between two fully collateralized parties B and C where B shorts the call option. The option is a liability of B's and is an asset to C. Under a full collateral Credit Support Annex (CSA, the legal collateral agreement used for swaps and other derivatives under ISDA Master Agreements), B has to post a cash amount same as the option value to party C, who has full and unrestricted control of the cash. In fact, even on day 1 when the option is transacted, C sends over $V_0$ cash amount to purchase the call option from B, and B has to post the same amount to C on the same day as day 1 margin. In the end, the same amount of cash comes back to C. C does not have any net cash flow, while having bought an ATM call option from B. C's treasury department does not need to come up with that cash amount for its option trader.

Same can be said of B: it doesn't see cash flow either while selling the call option. At the issue, therefore, a fully cash collateralized option has zero funding need for both counterparties.

Now on day 2, suppose the stock price increases and the ATM call becomes in-the-money, valued at $V_1$, $V_1 > V_0$. Under the CSA, B would have to shove over an amount of cash equal to $V_1 - V_0$ as variation margin (VM) to party C. Where does B get this cash amount? In other words, how does B fund the margin?

Since the option fair value equals to the bank account balance, $M_t = V_t$, we have $dV_t = dM_t$. But $dV$ is exactly the VM amount that needs to be posted. So the increment in the bank account is sufficient to cover the VM, so that party B does not need to arrange separately to fund the VM. Specifically, equation (4.b) suggests that the interest earned plus the net trading gain due to delta hedging would generate sufficient income that can be drawn to pay the VM.

Now if party B defaults at time s, with the option priced at V(s), party C already has V(s) amount of cash held as collateral and is not exposed to B's default event. In the end, we find that the trade is counterparty risk free and fully funded by the cash collateral account. The interest rate earned on the cash collateral is then the derivatives financing rate $r_d$ in equation (8.a & 8.b), with $r_d = r$.

Let $W_t$ denote the cash collateral account balance and $r_L$ denote the interest rate earned on the cash collateral, typically the Fed Fund rate in the US. $W_t$ enters the wealth equations (3), $\pi_t = M_t + \Delta_t S_t - V_t - L_t + W_t$. For full cash collateralized trades, we have $W_t = V_t$, so $M_t$ can be left to zero as $\pi_t$ is set to zero. This confirms that the cash collateral eliminates the need of a bank account. $r$ is replaced by $r_L$ in the financing equation (4.a and 4.b). Subsequently equation (7) also stands, but the appearance of $r$ will be replaced by the cash collateral rate $r_L$. So the discount rate now is the OIS rate, rather than an unspecified riskfree rate.

Some may say that the OIS rate is a good approximation of the riskfree rate, but we emphasize that we don't need to assume a riskfree rate at all. Because the OIS rates are marketwise rates and the assumed stock financing rates are also marketwise, accessible to every market participants, this rules out funding arbitrage by merit of equality among agents.

Cash collateral is a special case where the derivative financing rate $r_d$ is well defined to be the OIS rate and its discounting role well accepted (Piterbarg, 2010, and Hull and White 2013). It has the essence of a private derivatives financing market where the funding rate equals to the OIS rate. In other cases, we can also derive a discount rate applicable for different collateral asset types and margin schemes to determine an effective derivative financing rate.

## 3. Discounting when uncollateralized

At the opposite end of fully cash collateralized options is uncollateralized options. The buyer of the option initially funds the purchase. Subsequently, no cash/fund exchanges hands until the option expiry when the call option payoff needs to be paid. Or if one of the parties default before the expiry, a default settlement cash flow could result. As a contingent liability of the option seller, it is simply inviting to think that an uncollateralized or unsecured option should be discounted in the same way as the issuer's same ranked senior unsecured corporate debt does.

### 3.1. What is funding value adjustment (FVA)

End-users selling options to dealer-banks are quite familiar with credit valuation adjustment (CVA) charges. Take for example, suppose the option is written by a counterparty with a default probability of 5% per annum, the CVA is computed roughly at 2.6 so the final price is 29-2.6=26.4, where 29 is the default free option price, i.e., discounted at the OIS curve. It is not clear, however, what discount rate has been used to arrive at the final price of 26.4. Furthermore, CVA is computed at the derivative portfolio considered under a single netting set and allocated to each trade often based on simplistic rules. Such a CVA process necessarily blurs the discount curve further such that the final price is no longer associated with a discount rate as in equation (2).

For illustration purposes, let's assume this option is the only position the end-user has with the bank. The saved 2.6 is the price of buying a protection on the 29 current option price. Suppose the user actually spends that money to buy a protection from another bank D through a (contingent) credit default swap (CDS), paying exactly 2.6.

It is very important to know that protection afforded under a CDS contract is *unfunded*. In plain English, the fund to pay for the end-user's claims if party B defaults is not prepared or reserved beforehand. It is possible that the protection seller bank D might not be able to pay at then-time, for instance, if party B's default also triggers D's default. A (pre)funded protection is one that has set money aside in advance so that whenever B defaults, the user's default exposure is guaranteed to be fully covered. Funded protection has another significant operational advantage over unfunded protection in that it avoids the need of a CDS settlement, which in most cases has been shown to be a tedious process, which sometimes require lengthy legal intervention. The end-user would in no doubt favor the funded protection.

To obtain a funded protection, the end-user could short sell party B's corporate bond in a notional amount that matches his option exposure to B. If B defaults, the end-user will lose on his option exposure but gain on his short bond position. Since both bond and the option exposure are in the senior unsecured rank, they would recover at the same recovery rate and net out to no loss. The price to hedge is the manufactured dividend or par bond coupon, i.e., party B's senior unsecured debt interest rate. Because the bond rate is normally higher than the CDS premium, the user would demand a steep price cut from party B, i.e., higher than 2.6. Assume the bond rate is 5% per annum, then a rough estimate of the new adjustment amount will be 4.33, making a new final price of 29-4.33 = 24.67. The difference between this new adjustment of 4.33 and the old CVA of 2.6 is an adjustment responsible for the funded vs unfunded protection difference, and is thus called funding value adjustment (FVA).

Dealers and banks have come up with slightly different perspectives. FVA was initially introduced to take into account of the fact that banks can't fund their derivatives operations at the LIBOR level and as a result banks' treasury departments charge derivatives trading desks internal funding rates aligned with their secondary market bond yields. Specifically, consider dealer B enters a long option with an uncollateralized client C and hedges (shorts) the option with another dealer bank A under a full CSA. Because party C does not post collateral and B has to post cash collateral to A, the trader has to borrow the cash collateral from his internal treasury department. If his treasury charges an interest rate of $r_b$, then his net cost is $r_b - r_L$ where $r_L$ is the interest rate earned on cash collateral. The net interest cost applies to the funding amount of $V^*$, the option fair price under a full CSA, i.e., OIS discounted, default riskfree option price. By discounting each small time period's cost $(r_b - r)V^* ds$ and summing them up, traders define FVA for uncollateralized derivatives as follows,

$$X = E_t[\int_t^T (r_b - r)V^*(s)D(t,s)ds] \qquad (9)$$

where $D(t,s)$ is an applicable and debatable discount factor from time $s$ to time $t$, $s>t$.

For some, it is the product of the riskfree discount factor and B's own survival probability, capturing the fact that such a cost is only incurred while party B itself is operating. Others argue that it should be the product of the riskfree discount factor and the joint survival probability of party B and his counterparty C, for the collateral funding cost also disappears once party C defaults and the short option settles. Regardless which way it goes, there is no clear identification of a discounting curve for the final price $V=V^*-X$, as $V^*$ is discounted at the OIS rate and $X$ is something else.

This form of FVA has been controversial, invalidated in theory but nonetheless implemented in practice. For example, JP Morgan's Q1 2014 report announced 1.5 billion dollar loss due to FVA charge on OTC derivatives and structured notes, citing as an illustration 50 bp FVA funding cost on 5 year duration, see Whittall (2014). Anderson, Duffie and Song (2019) prove that this kind of accounting loss is not justified and FVA as a cost of acquiring a derivative asset should land in the firm's equity rather than its balance sheet. The most prominent argument (Hull and White 2014) has been that FVA calculations such as (9) could result in multiple fair values which depend on the pricing agents own funding curves and violate the law of one price, a topic of debate yet to close as theory and practice simply don't seem to be converging.

### 3.2. Liability-side pricing principle

When a party, say B, carries an OTC derivative asset uncollateralized, it has an unsecured, senior ranked exposure to its counterparty -- party C. Because its primary risk can be hedged by dynamic trading in the underlying stock, the remaining risk is really credit risk, which resembles the usual corporate credit risk. Lou (2015b, 2016) provides a proof and generalizes it as the liability-side pricing principle, which states that an uncollateralized derivative's future cash flows should be locally discounted at liability-side's senior unsecured rate. For a swap or swap portfolio, the instantaneous discount rate switches between party B and C's local rates, depending on whether the forward npv is positive (an asset) or negative (a liability).

The key idea is that the *fair* instrument, for the parties of a bilateral derivative trade to hedge their counterparty exposure and to fund the derivative, is to have the party on the liability side deposit cash with the party on the asset side. The party making deposit is economically neutral if the deposit will earn a coupon rate same as the market rate of its debt. The deposit made outside of the derivative's netting set would offset the derivative exposure via the Set-off provision in the ISDA Master Agreement or common laws. Equivalently, if the party's bond is traded in a liquid market, a short hedge achieves the same purpose.

To demonstrate, we consider a one-step binomial tree model shown in Figure 1. The stock price starts at 50, then either goes up to 64.201 at a probability of 0.4652, or goes down to 38.940. The option is issued by party C which could default during the period. Suppose that at $t_0$, party B knows the price of the option at 6.468 and pays C that amount to purchase this uncollateralized option. Holding on the option after the purchase, B has an unsecured exposure to party C.

|  | step 0 |  | step 1 |  |
|---|---|---|---|---|
|  |  |  | 64.201 | stock prc |
|  |  |  | 14.201 | opt payoff |
|  |  |  | 6.607 | deposit |
| stock prc | 50.000 | C default |  |  |
| option val | 6.468 | set-off, no cash flow |  |  |
| C's deposit | 6.468 |  |  |  |
|  |  |  | 38.940 | stock prc |
|  |  |  | 0.000 | opt payoff |
|  |  |  | 6.607 | deposit |

Figure 1. One-step binomial tree ($dt=0.25$ year) illustrating the liability-side pricing principle.

To hedge its exposure to C, B could short C's senior unsecured bond. Or as Lou (2015b) suggests, B could ask C to deposit $6.468 cash on which B would pay 8.5% interest rate, same as C's par bond coupon rate. C is economically indifferent as he could raise the cash from the bond market at the same rate. The new liability assumed by issuing in the market place offsets his deposit (on the asset side) so that there is no net balance sheet impact. From cash flow perspective, C receives the market rate from B and pays out the same to bond investors. No cash flow is generated.

From party B's perspective, the derivative asset is financed by C's deposit. No other forms of financing, including from own treasury, is necessary. Now imagine if C defaults, B has two transactions with C: the derivative receivable, and the deposit as a payable. Under common laws, the right to set-off would apply to determine the final settlement amount. Since the deposit amount is same as the derivative exposure, the derivative contract and the deposit set off completely. The net balance is zero and there is no default settlement cash flow. The cash deposit serves both purposes of hedging counterparty credit exposure and financing the derivative. The total economic cost of hedging and funding is simply the market rate of C's debt[5].

The question then becomes how B will fund this 8.5% par coupon. The liability-side pricing model (Lou, 2015b, 2016) puts this squarely back to the option pricing theory where the option has to accrue and to be discounted at this higher, non-riskfree rate. The B-S equation is revised as follows,

$$\frac{\partial V}{\partial t}+(r_s-q)S\frac{\partial V}{\partial S}+\tfrac{1}{2}\sigma^2 S^2\frac{\partial^2 V}{\partial S^2}-r_e V=0, \tag{10.a}$$

---

[5] Here the derivative asset attracts liquidity at C's unsecured rate. Thus $r_c$ is the derivative financing rate. But this is theoretical or implied, as holding an uncollateralized asset does not actually generate liquidity.

$$r_e(t) = r_b I(V(t) \leq 0) + r_c I(V(t) > 0) \qquad (10.b)$$

where $r_b(t)$ and $r_c(t)$ are B's and C's bond interest rates respectively, $I(.)$ is an indicator function, and $r_e(t)$ is the effective discount rate. Equation (10) is written from party B's perspective, i.e., positive $V$ means that the option is an asset to B and a liability to C and C is said to be on the liability-side. If we split $V$ into a positive part $V^+$ and a negative part $V^-$, so that $V = V^+ - V^-$, then $r_e(t)V(t) = r_c V^+(t) - r_b V^-(t)$. The asset or receivable part of the derivative is discounted at the counterparty's funding rate, while the liability or payable part at the firm's own rate. Regardless, the discount rate is always on the liability-side.

Accordingly, the risk neutral pricing formula under $Q$-measure for an option with a terminal payoff function $H(T)$ is extended,

$$V(t) = E_t^Q [e^{-\int_t^T r_e(u)du} H(T)]. \qquad (11)$$

In the example given in Figure 1, B holds the option as an asset. C is the option writer or issuer, so $r_e = r_c$, that is C's rate of 8.5% becomes the discount rate.

The (total) counterparty risk adjustment (CRA), denoted by $X$ as the difference between the riskfree derivative price $V^*$ and $V$, is precisely given by,

$$X = V^* - V = E_t [\int_t^T (r_e - r) V^*(s) e^{-\int_t^s r_e du} ds] \qquad (12)$$

This formula is interesting for it shows $r_e$ as both the discount rate -- appearing in the exponent, and the financing rate -- appearing in the funding spread applied to $V^*$.

Until today, many derivatives end-users have remained performing LIBOR discounting for uncollateralized trades, although Hull and White (2013) argue that they should be discounted at the same OIS rate as for collateralized trades, but formula (11) shows neither role for LIBOR nor for OIS. Pre-crisis LIBOR rates are thought to roughly reflect the credit risk profile of typical 'AA' rated banks who were able to fund in the proximity of LIBOR rates (Huge and Lando 1999). So to some, LIBOR rates are thought to be a good approximation to the riskfree rate as they had been mostly a few basis points away from other riskfree rate proxies such as the Treasuries repo rates. The industry's adoption of the LIBOR curve, however, is based on their funding experience (Hull and White 2013) that the LIBOR swap curve is closer to where they fund OTC derivatives operations. Thanks to the financial crisis, we have finally come to a better understanding of how to discount the uncollateralized derivatives.

Discounting unsecured exposures at the issuer's unsecured rate, while seemingly as natural as discounting an issuer's bond at the issuer's yield curve, has proven to be evasive in bilateral defaultable OTC derivatives pricing. In fact, if one takes a corporate bond's cash flows and prices it with the standard reduced form credit model calibrated to the CDS market, a different yield would result. The market has come to acknowledge the difference in CDS implied bond yield and market yield as the basis. Without incorporating the basis, the reduced form model is unable to reprice a zero coupon bond at the market, which is trivial shown in equation (11). In this sense, the liability-side pricing of uncollateralized derivatives is consistent with bond market pricing, as it correctly recovers the natural discount rate.

A side-effect of the liability-side discounting is that it doesn't matter whether the base index is LIBOR or its replacement – SOFR. For cash collateralized derivatives, we use OIS curves; for uncollateralized, we apply parties' bond curves.

**3.3. Partial collateralization**

In a bit more general setting where the derivative is only partially cash collateralized and partially uncollateralized, the B-S equation can be shown to become

$$\frac{\partial V}{\partial t} + (r_s - q)S\frac{\partial V}{\partial S} + \tfrac{1}{2}\sigma^2 S^2 \frac{\partial^2 V}{\partial S^2} + r_b(V-L)^- - r_c(V-L)^+ - r_L L = 0, \quad (13)$$

where $L_t$ denotes the cash collateral amount, $|L|<|V|$, $V$-$L$ the uncollateralized amount, and $r_L$ earned interest rate on $L_t$.

If we write,

$$\eta_c = \frac{L^+}{V^+}, \eta_b = \frac{L^-}{V^-}$$

then PDE (13) becomes equation (10.a) with the effective rate (10.b) replaced by,

$$r_e = (r_c(1-\eta_c) + \eta_c r_L)I(V>0) + (r_b(1-\eta_b) + \eta_b r_L)I(V \leq 0) \quad (14)$$

The effective discount rate is an average of the cash collateral rate and the unsecured rate with weights reflecting the proportion of secured and unsecured exposures.

## 4. Liquidity rate for segregated cash collateral

It is worth noting that in the derivation leading to PDEs and the effective discount rates (10) and (14), the cash collateral is allowed to comingle with the economy, typical for variation margin of the derivatives' mark-to-market. Small volume end-users can elect to have their collateral segregated. Comingled cash is usable for the economy's general operations, and economically serves both purposes of credit mitigation and derivatives funding. Cash collateral is imperfect when segregated, as credit protection is provided at the cash amount, but no part of it can be used to fund the derivatives. This becomes an intermediate case between (no-segregated) cash collateral that provides both protection and funding, and uncollateralized case that offers neither protection nor funding.

To determine the applicable discount rate, let's consider party B and party C enter into an option trade under a CSA where full cash collateral is required but segregated in a separate bank account. Collateral posting mechanism is exactly same as the comingled case, i.e., both day 1 and variation margins are posted, except the posted cash is deposited in the bank account administrated by a trustee. Any interest earned on the bank account is promptly returned to the pledger by the trustee and is out of the benefactor's reach. Should a credit event happen, the deposit account will be drawn to pay for default settlement.

Because party B does not have access to the day 1 margin posted, B then has to fund the option purchase. Now assume party B has a financing account in $N_t$ with rate $r_n$, then $N_0$ equals to the purchase price. The only difference of this segregated economy from the non-segregated economy of section 1.3 is default settlement. At the time of C's default, cash will be released from the segregated bank account to B to cover any close-out payment.

Following the same derivation procedure as in Section 1.1, with the funding account $N_t$ replacing the money market account to provide fund at a rate of $r_n$ instead of $r$, we can easily obtain the same PDE as equation (10) except the discount rate is replaced by $r_n$.

A quick and seemingly natural choice of $r_n$ is bank B's own funding rate. Such a choice would lead to the problem of having multiple fair values when multiple banks start to price it, in violation of the law of one price. A rational choice has to consider the economic content of the segregated collateral. To illustrate, B can segregate the option (as an asset) into a trust to issue a credit linked note (CLN) referencing the option and its segregated cash collateral. The CLN is linked to C's credit, meaning that when C defaults, it will liquidate its reference assets and pay the noteholder. The issued amount of the CLN equals to the amount of segregated cash collateral, $N_t = L_t$. CLN is priced at par with $r_n$ being its rate. The CLN noteholder can be fully repaid upon default termination, when the segregated cash collateral is released. So the CLN is subject to default event but won't incur losses. Its pricing rate $r_n$ should therefore not contain C's default risk. It reflects bond market liquidity and other non-credit, market structural factors (Hull and White 2014), and can be referred to as the issuer liquidity rate, denoted by $\mu$.

A firm's liquidity rate might not be directly market observable but it can be inferred from the cash bond market and CDS market via a basis trade. Suppose a default riskfree party sets up a basis trade by holding a par bond at (continuously compounded) coupon rate $r_c$ issued by C and buying protection via a CDS on the issuer of the same maturity. Assuming that CDS is cleared through a CCP and is continuously rolled at CDS premium $x$, which is free of counterparty risk and also continuously compounded. Without considering gap risk and bond and CDS market liquidity risk, the basis trade is not subject to C's default risk, so its no-arbitrage pricing rate has to be $r_c - x$, i.e., $\mu_c = r_c - x$. $\mu_c - r = r_c - x - r$ is then the bond CDS basis (or funding basis).

As an example, a firm may have a credit spread of 70 bps, i.e., $r_c - r = 0.7\%$, while only 20 bps of it attributes to the liquidity basis, $\mu_c - r = 0.2\%$, so 50 bps is due to default risk, $r_c - \mu_c = 0.5\%$. Under zero-recovery assumption, $\mu_c = r_c - \lambda_c$, where $\lambda_c$ is C's default intensity.

The liquidity rate's fit for fair value purposes can also be illustrated by an informal market equilibrium argument. Because $r_n$ directly impacts derivatives pricing, party C has vested economic interest in $r_n$. If $r_n$ is too high, for example, and C finds his own funding cost is lower, C would step in to buy the CLN, willing to pay at his own funding liquidity rate, thus lowering $r_n$. If $r_n$ is already low in the market place, C benefits, but B would want to have a claim of that benefit, essentially driving up $r_n$ back to C's liquidity rate in the equilibrium. So in the end, $r_n = \mu_c$.

Depending on the sign of the derivative fair value, $r_n$ will be switching as well, and could be replaced by

$$r_e = \mu_c I(V > 0) + \mu_b I(V \leq 0). \tag{15}$$

The derivative is still financed on the liability-side, with a cost equal to liquidity basis or the bond-CDS basis, $\mu_b - r$ or $\mu_c - r$. PDE (10) with $r_e$ in equation (15) remains the law of one price conforming.

In the case where the segregated amount $L$ is only a fraction of V (with the rest being unsegregated cash collateral), for $V \neq 0$,

$$\eta_c = \frac{L^+}{V^+}, \eta_b = \frac{L^-}{V^-} \tag{16}$$

The discount rate can be shown to be

$$r_e = [r_c(1-\eta_c) + \eta_c \mu_c] I(V > 0) + [r_b(1-\eta_b) + \eta_b \mu_b] I(V \leq 0) \tag{17}$$

More generally, we could introduce a flag $\chi$, 0 meaning segregated, 1 comingled. Since non-segregated cash returns at rate $r_L$, the effective discount rate for the PDE can be rewritten to accommodate both cases,

$$\begin{aligned} r_e &= r_{ec} I(V > 0) + r_{eb} I(V \leq 0), \\ r_{eb} &= r_b(1-\eta_b) + \eta_b[(1-\chi_b)\mu_b + \chi_b r_L], \\ r_{ec} &= r_c(1-\eta_c) + \eta_c[(1-\chi_c)\mu_c + \chi_c r_L] \end{aligned} \tag{18}$$

For party C, its effective rate $r_{ec}$ is a linear combination of the unsecured rate $r_c$, the liquidity rate $\mu_c$, and the cash rate $r_L$, each applied to the unsecured (and unfunded) portion 1-$\eta_c$, secured but unfunded portion (1- $\chi_c$)$\eta_c$, and $\chi_c\eta_c$ the remaining secured and funded portion of the fair value V⁺. $r_{eb}$ can be understood similarly.

This concept of an effective derivative financing rate in the presence of counterparty credit risk and collateral is also evident in Hull and White (2014), where $r_d$ is shown using dealer or average dealers' rates rather than liability-side's rates. It is not seen in Burgard and Kjaer (2011), although their main result 1 (when the close-out uses risky market value) can be easily converted to yield $r_{ec} = (1-R_c)(r_c - r) + r_F$ and $r_{eb} = (1-R_b)(r_b - r) + r$, where $R$ is the recovery rate, $r_F$ equals $r_{eb}$ since the derivative can't be repo-ed. But these are still different from (18). Brigo et al (2014) and Anderson, Duffie, and Song (2019) do not present such a form, as they follow the reduced form risk neutral pricing framework.

When securities are used as derivatives collateral, their quantities are determined by collateral haircuts – discounts applied to their market values. And they have to go through the repo market to transform into cash, and are subject to repo haircuts. The differences in collateral haircuts and repo haircuts create pockets of unsecured exposure or secured but segregated exposure. The resulting discount rate is found to be a mixture of the OIS rate, the senior unsecured rates, and the liquidity rates of the counterparties, see Lou (2017).

## 5. Binomial Tree Pricing

The switching effective discount rate in equation (10.b, 14 and 18) depends on the sign of the fair value, so the fair value formula in (11) is generally recursive, unless the sign of the derivatives is known for sure such as a call option. This type of coupling between the discount rate and the fair value reflects the nonlinear nature of pricing with counterparty risk. In our construct, the only nonlinearity appears in the effective discount rate, unlike Burgard and Kjaer (2011), where the coupling binds the unknown fair value and its adjustment through exposure amounts. Brigo et al (2014), for example, have to introduce a separate non-linear valuation adjustment in their backward stochastic differential equation (SDE) to correct an overlap in value adjustments.

Synthesizing derivative collateralization effect into one derivative financing rate has its advantages. For example, it becomes obvious that party C's zero coupon bond will be priced at C's senior unsecured rate $r_c$, a critical model consistency test not satisfied by all. Moreover, it allows us to bypass valuation adjustments to compute the fair value directly.

This coupling could post a challenge to numerical solutions. For low dimension pricing problems such as an equity derivative portfolio with a single underlying or a swap portfolio on the same reference rate (e.g., LIBOR), a finite difference method is developed (Lou 2015b). Also as demonstrated in Lou (2016), a Monte Carlo simulation with regression procedure can be developed focusing on the convergence of the discount rate. In this section, we enlist a binomial tree procedure to help understand the role of the effective discount rate and to provide a rudimentary implementation.

Consider a one-step tree where the stock price starts at S, goes up to $S_u$, or down to $S_d$. The option payoff at up and down nodes are $H_u$ and $H_d$ respectively. At the moment, we don't know what discount rate, $r_b$ or $r_c$, to use yet as it depends on the option fair value at $t_0$. One could use $r_b$ to discount the expected payoff of $pH_u+(1-p)H_d$ to arrive at a value $v_b$. If $v_b$ is negative, then we know the choice of $r_b$ is correct, so $V= v_b$. Otherwise, we should do it again with $r_c$ as the discount rate. This kind of trial and error is not necessary, however. If we write down the one step pricing formula,

$$r_e = r_c I(V > 0) + r_b I(V \leq 0),$$
$$V = [pH_u + (1-p)H_d]e^{-r_e T} \tag{19}$$

then since the discount factor $e^{-r_e T}$ is always positive, the sign of $V$ is the same as the expected payoff $pH_u + (1-p)H_d$, i.e. $I(V > 0) = I(pH_u + (1-p)H_d > 0)$. We now have the following decoupled discounting and pricing formula,

$$r_e = r_c I(pH_u + (1-p)H_d > 0) + r_b I(pH_u + (1-p)H_d \leq 0),$$
$$V = [pH_u + (1-p)H_d]e^{-r_e T} \tag{20}$$

Below we give an example on a 2-step binomial tree used to price an uncollateralized option strategy which has mixed positive and negative terminal values. A shifted stock forward is a combination of a long European call option at a lower strike $K_1$ and a short put at a higher strike $K_2$ of the same expiry. It has a positive payoff when the terminal stock price $S_T$ is greater than the mean of $K_1$ and $K_2$ and negative payoff otherwise. Figure 2 shows a sample binomial tree solving the forward's fair value with $K_1 = 45$ and $K_2 = 55$.

The unsecured rates for bank B and customer C are 5.7% and 8.5% respectively. Notice that at step 1 up node, it is an asset to B, so its counterparty C's rate is used to discount. At the lower node, *npv* is negative, so it becomes a liability of B and B's own rate of 5.7% is used. The final price is 0.955. Note the same tree can be used to compute the default risk free price by simply replacing the discount rates with the OIS rate. For those interested in valuation adjustments, CRA defined in (12) as the difference between the riskfree price and the final price is 0.066. Option Greeks such as delta, gamma and vega can be obtained in the usual way.

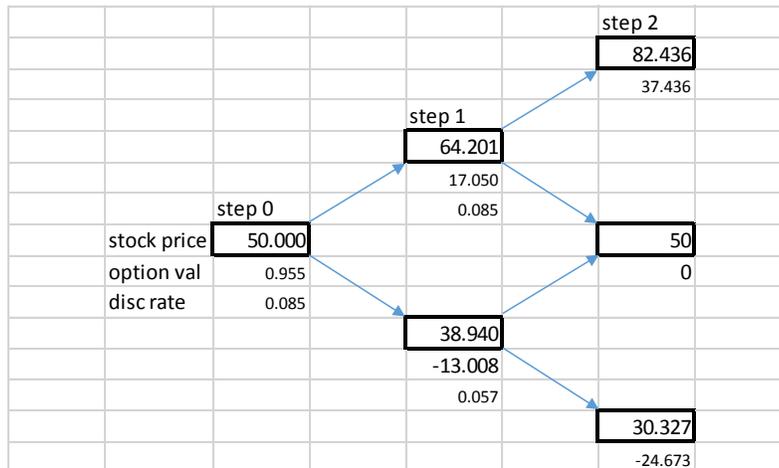

Figure 2. Two step binomial tree computing the fair value of an uncollateralized option portfolio consisted of a long call at 45 strike and a short put at 55 strike, i.e., a synthetic shifted forward, with expiry in 6 months, repo rate 5.5%, volatility 50%, up probability 0.4652.

This simple example can be easily extended to handle American options, and a portfolio of options on the same underlying stock. Similarly, an interest rate tree can be built to price swaps and interest rate derivatives between two defaultable counterparties. Unlike derivatives portfolios or netting sets between dealer-banks which involve many different underlying instruments, end-users' derivatives portfolios are often segregated or one-sided which may find the binomial tree procedure helpful.

**6. Value adjustments (XVA) based on decomposition of the discount rate**

From the pricing point of view, once the discount rate is determined, we could proceed to solve for the final price and its sensitivities or *Greeks* as desired. XVAs are only necessary because current models typically separate pricing of primary market risk factors (such as stock price, volatility, and interest rate) and counterparty risk factors. When separately computed, XVA often are not consistent with the final price. It is pointed out, for example, that practitioners' FVA leaves out interest rate sensitivity unhedged, therefore adding FVA to the riskfree price does not yield the true fair price (Lou 2016).

In Lou (2015b), the total counterparty risk adjustment (CRA, equation 12) of an uncollateralized trade is decomposed into a coherent form of bilateral CVA and FVA, which corresponds to a decomposition of the spread $r_e-r$ into a default risk measure and the funding or liquidity basis. This form of definition avoids overlapping DVA (debt value adjustment, a benefit related to not having to pay the full amount of a liability after a default event of the issuer) of a derivatives liability and its FVA benefit (of not having to fund one's own derivatives liability), is balance sheet neutral, conforms to the law of one price, and is thus better suited for fair value purposes.

This decomposition is self-consistent and avoids any overlap among *XVA*s, in that adding up all XVAs results in the total adjustment which is precisely defined. To compare with typical XVA definitions, below we list our version of CVA and FVA for a pure asset such as an uncollateralized option,

$$cva = E_t[\int_t^T (r_c - \mu_c)V_s^* e^{-\int_t^s r_c du} ds] \tag{21}$$

$$fva = E_t[\int_t^T (\mu_c - r)V_s^* e^{-\int_t^s r_c du} ds] \tag{22}$$

Note the discount rate in the above is always the unsecured rate of the issuer C. Typical industry unilateral CVA definition is written as

$$CVA = E_t[\int_t^T (1-R)V_s^* e^{-\int_t^s r du} Q(s)\lambda ds] \tag{23}$$

where $R$ is the recovery rate of the counterparty, $\lambda$ is C's default intensity or hazard rate, $Q(s)$ could be the survival probability of party C or the joint survival probability of B and C. $Q(s)\lambda ds$ is the (risk-neutral) default probability in time period *[s, s+ds]*. Since we normally hold $(1-R)\lambda = (r_c - \mu_c)$ true, the difference between (21) and (23) is that (21) uses C's bond discount while (23) uses the product of riskfree discount and the survival probability. The problem with (23) is that even if we define FVA solely on the liability-side's funding basis in accordance with (23), i.e., write FVA as

$$FVA = E_t[\int_t^T (\mu_c - r)V_s^* e^{-\int_t^s rdu} Q(s)\lambda ds] \qquad (24)$$

CVA (23) plus FVA (24) does not return to the total valuation adjustment in equation (12) and will not reprice a simple debt instrument such as a bond.

Anderson, Duffie and Song (2019) prove that traders may incorporate the FVA cost into bid/ask to recoup value for equity holders, but not the fair value, although it remains subject to a CVA adjustment to account for the probability of counterparty default. Their formula doesn't seem to reprice a bond and they have been silent about the basis. It could be interesting to see how Anderson et al (2016)'s framework could be extended to incorporating the basis.

Our *FVA* (eqt. 22) corresponds to the liquidity component of funding an uncollateralized exposure. This is consistent with Hull and White (2014)'s finding that a dealer bank's funding of an uncollateralized asset would incur a debt valuation adjustment (DVA) that offsets its FVA cost to the extent that only the dealer's basis contribution is left. Our version of FVA however is attributed to the bond-CDS basis of the issuer, not to that of the bank, and, as a result, is compliant of the law of one price.

Note that our *XVAs* use the same discount rate as the pricing equation, which facilitates numerical solutions. The binomial tree model developed in Section 4 can also be extended to compute these various adjustments, taking advantage of the switching rates being the same for XVA as for the fair value. To facilitate binomial tree computation, cva formula can be rewritten,

$$cva_{t1} = E[\int_{t1}^{t2}(r_c - \mu_c)(1-\eta_c)I(V>0)V_s^* e^{-\int_{t1}^{s} r_e du} ds + e^{-\int_{t1}^{t2} r_e du} cva_{t2}] \quad (25)$$

where $cva_{t2}$ is the cva on a node at time $t_2$, and we compute cva backward from $t_2$ to $t_1$, $t_2 > t_1$. At expiry $T$, $cva_T = 0$. Other XVAs are similar.

The total counterparty risk value adjustment (CRA) can be broken down into bilateral CVA composed of cva and *dva* with *CVA=cva-dva*, and bilateral FVA composed of *cfa* and *dfa* with *FVA=cfa-dfa*, per Table 1 below,

Table 1. XVA of long or short shifted forward breaks down into CVA (cva and dva) and FVA (cfa and dfa). The riskfree price is 1.09. The unsecured rates for bank B and customer C are 5.7% and 8.5% respectively. B and C's liquidity rates are 5.2% and 5.5%, leaving out 0.5% and 3% as credit default rates respectively.

|           | Short Fwd | Long Fwd |
|-----------|-----------|----------|
| final prc | 1.032     | -1.127   |
| Cva       | 0.059     | 0.044    |
| Cfa       | 0.01      | 0.007    |
| Dva       | 0.008     | 0.01     |
| Dfa       | 0.003     | 0.004    |
| CRA       | 0.058     | 0.037    |

Figure 3 shows how such a computation is done for the case of cva. Each node's option fair value and discount rate are same as in Figure 2. The blacked box now shows the riskfree fair value to facilitate the XVA calculations.

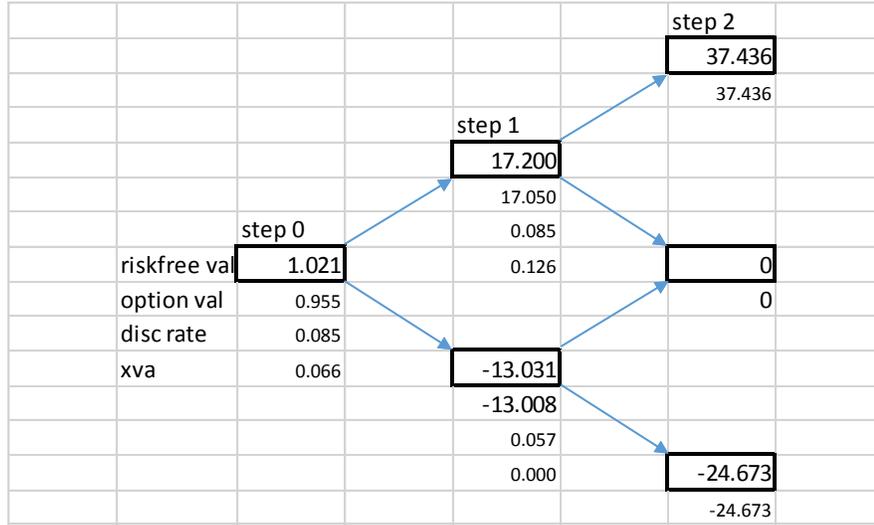

Figure 3. Using the same binomial tree to compute XVA. Shown example is for cva.

One way to break out CRA (Lou 2015b) is to run the binomial tree with both parties using the OIS discount rate to get $V^*$, then shift party C's discount rate to its liquidity rate $\mu_c$ to get $V_1$. Then cfa (credit funding adjustment, part of FVA due to counterparty) equals to $V^*-V_1$. Further shifting C's discount rate to its unsecured rate $r_c$ to compute $V_2$, then cva equals to $V_1-V_2$. Now keeping C's discount rate as is, and shifting B's rate to its liquidity rate $\mu_b$ to compute $V_3$, then the self-funding benefit, dfa equals to $V_2-V_3$. Lastly, shifting B's rate to its senior unsecured rate $r_b$ to get $V$, and dva = $V_3-V$. This is different from Figure 3, where they are computed using the same effective discount rate throughout, but the total valuation adjustment is conserved.

The binomial tree model can be easily built for interest rates or foreign exchange (FX) rates so that swap and FX derivatives portfolios can be evaluated by end-users. Admittedly for higher dimension problems, for example, a derivatives portfolio involving many stocks, the binomial tree approach is of limited use. Such is the case, the Monte Carlo (MC) simulation approach presented in Lou (2016) can be utilized to compute XVA.

## 6. Conclusions

The risk-neutral pricing formula where derivatives' cash flow is discounted at the riskfree rate to arrive at the fair value has long drifted away from us. In particular, after the financial crisis of 2007-2009, various value adjustments proposed in a variety of ways have obscured this mathematically beautiful and financially intuitive formula. Derivatives pricing and valuation have become overly complex and almost inaccessible to end-users in the investment community. This paper attempts to recover the formula by generalizing a new discount rate, to take into account of counterparty credit risk, funding, margin and collateralization.

At one extreme is the well accepted OIS rate, applicable to fully cash collateralized derivatives. At the other extreme is the issuer's senior unsecured borrowing rate, applicable to uncollateralized derivatives. For derivatives collateralized with segregated cash, the proper discount rate for the fair value is identified as the issuer's liquidity rate.

Presentation of such a rate extends the classic risk neutral pricing formula from default free counterparties to bilateral defaultable counterparties. The risk neutral measure stays and the payoffs are the same: only the riskfree rate needs to be replaced to arrive at the final price that reflects counterparty risk and its mitigation effect by means of margin and collateralization. Although the effective discount rate is coupled with the fair value in general, a binomial tree could unlock the coupling. In particular, an end-user friendly binomial tree model enables the final price to be computed efficiently, useful for low dimension portfolios such as interest rate swaps, foreign exchange swaps, and individual equity or commodity derivative portfolios.

Our discount rate formulation enables derivatives end-users to dissect various forms of adjustments, or XVAs, by decomposing the effective derivative financing rate into relevant risk factors. CVA (C = credit), for example, corresponds to the default risk as usual, and FVA (F = funding) to the funding risk component of the bond credit spread. Contrary to current industry implementations where the default free fair value is computed in one system and XVAs are computed in other systems, sometimes with a number of distinct XVA models, this allows XVAs and the fair value to be computed simultaneously within one module.